# ReaLPrune: ReRAM Crossbar-aware Lottery Ticket Pruned CNNs

Biresh Kumar Joardar, *Member, IEEE,* Janardhan Rao Doppa, *Member, IEEE,* Hai (Helen) Li, *Fellow, IEEE,* Krishnendu Chakrabarty, *Fellow, IEEE,* and Partha Pratim Pande, *Fellow, IEEE*

**Abstract**—Training machine learning (ML) models at the edge (on-chip training on end user devices) can address many pressing challenges including data privacy/security, increase the accessibility of ML applications to different parts of the world by reducing the dependence on the communication fabric and the cloud infrastructure, and meet the real-time requirements of AR/VR applications. However, existing edge platforms do not have sufficient computing capabilities to support complex ML tasks such as training large CNNs. ReRAM-based architectures offer high-performance yet energy efficient computing platforms for on-chip CNN training/inferencing. However, ReRAM-based architectures are not scalable with the size of the CNN. Larger CNNs have more weights, which requires more ReRAM cells that cannot be integrated in a single chip. Moreover, training larger CNNs on-chip will require higher power, which cannot be afforded by these smaller devices. Pruning is an effective way to solve this problem. However, existing pruning techniques are either targeted for inferencing only, or they are not crossbar-aware. This leads to sub-optimal hardware savings and performance benefits for CNN training on ReRAM-based architectures. In this paper, we address this problem by proposing a novel crossbar-aware pruning strategy, referred as ReaLPrune, which can prune more than 90% of CNN weights. The pruned model can be trained from scratch without any accuracy loss. Experimental results indicate that ReaLPrune reduces hardware requirements by 77.2% and accelerates CNN training by ~20× compared to unpruned CNNs. ReaLPrune also outperforms other crossbar-aware pruning techniques in terms of both performance and hardware savings. In addition, ReaLPrune is equally effective for diverse datasets and more complex CNNs.

✦

## 1 INTRODUCTION

DEEP learning (DL) has enabled significant growth in diverse real-world applications ranging from face recognition and AR/VR to natural language processing and computer vision [1]. However, existing DL applications are computation intensive and are typically deployed on the cloud. For example, most voice assistants, e.g., Apple Siri and Microsoft's Cortana, are based on cloud computing and do not function if the network is unavailable. Smaller edge devices are insufficient to support the computations associated with many DL algorithms including the training of Convolutional Neural Networks (CNNs). However, there is a growing necessity to address the problem of training on the edge platforms due to a multitude of factors [1][2][3]. First, for many important applications (e.g., mobile health and recommendation systems), privacy and security are important concerns [4][5]. Second, vendors often want to personalize their ML-driven products/applications for each user; this will require incremental learning on the edge. For instance, Google G-board uses federated learning to collaboratively train the typing prediction model on smartphones [3]. Each user uses their own typing records to train G-board. Hence, the trained G-board can be used immediately, powering experiences that is personalized to each user. Third, there is a need to increase the accessibility of AI applications to different parts of the world by reducing the dependence on the communication fabric and the cloud infrastructure. Fourth, many of these applications such as robotics and AR/VR require low latency, which may not be achievable without performing the computation directly on the edge platform. However, existing edge platforms are still not capable of training large ML models such as CNNs and addressing this problem is one of the primary foci of this work.

Emerging resistive random-access memory (ReRAM) has been shown to be an effective platform for efficient training and inferencing of deep learning algorithms, including CNNs [6][7]. ReRAM-based systems can be used to enable low-power training on the edge. Recently ReRAM-based prototypes for CNN inferencing and training have been demonstrated [8][9]. ReRAM crossbars can efficiently perform matrix-vector multiplication, which forms the backbone of most CNN computations [7]. Prior work, such as Pipelayer [6] and AccuReD [10], have shown that ReRAM- based architectures can outperform GPUs for training CNNs while consuming less energy. In addition, ReRAM-based systems are more area-efficient compared to their GPU counterparts and do not require expensive off-chip memory access due to their "in-memory" nature of computation [7].

Despite these advantages, ReRAM-based architectures are not scalable with the size of CNNs. Deep CNNs (i.e., CNNs with many layers) involve many weights, which necessitates many ReRAM crossbars for storage and

- *B. K. Joardar, H. Li, and K. Chakrabarty are with the Department of Electrical and Computer Engineering, Duke University, Durham, NC 27708; E-mails: {bireshkumar.joardar, hai.li, krish}@duke.edu;.*
- *J. R. Doppa, and P. P. Pande are with the School of Electrical Engineering & Computer Science, Washington State University, Pullman, WA 99163; E-mails: {jana.doppa, pande}@wsu.edu.*

computation [10]. Unlike GPUs where each CNN layer is processed one after another, ReRAM processes the CNN layers in a pipelined fashion [6]. In a pipelined implementation, all layers of the CNN are active at the same time, i.e., the computations of all the layers are performed simultaneously; each layer processes a different input image. Hence, ReRAM-based architectures must store all the weights on-chip, which necessitates many crossbars. Moreover, training requires storing intermediate data (such as activations) to be used during the backward phase of training [6]. Overall, this necessitates many ReRAM crossbars for storage and computation. This is especially problematic for deep CNNs such as VGG-19. VGG19 involves 143 million weights and storing all the weights will necessitate at least 143 million ReRAM cells, which is expensive in terms of both area and power, especially for edge devices. Clearly, this problem must be addressed to enable training on end user devices.

Pruning is an effective way to reduce the amount of storage and computation needed for CNN training/inferencing [11]. Pruning reduces the number of weights in a CNN by forcing some of the weights to be zero. Multiplications and additions with zero are functionally redundant; multiplication with zero yields a zero and the sum of any number with zero is the number itself. Hence, we can safely omit storing and computing with zero weights. This can lead to potential savings in terms the total number of ReRAM crossbars necessary for CNN training. However, conventional pruning techniques are oblivious to the crossbar structure. Simply pruning weights does not translate to ReRAM crossbar savings without the knowledge of the underlying mapping mechanism. Crossbar-aware pruning strategies can solve this problem. Crossbar-aware pruning approach results in relatively more hardware savings despite potentially lower levels of sparsity than their conventional ReRAM-unaware counterparts. However, existing crossbar- aware pruning strategies are targeted for CNN inferencing only [12][13]. The pruned networks obtained using these methods cannot be trained from scratch without accuracy loss [11]. Moreover, CNN training requires the storage of both weights and activations. As a result, simply pruning the weights may not lead to significant reductions in the number of activations that must also be stored during training.

Therefore, there is a clear need to develop new crossbar-aware pruning strategies for CNN training on smaller devices. Towards this goal, we propose the first ReRAM crossbar-aware pruning technique for CNN training, which aims to reduce both the number of weights and activations that must be stored on ReRAM crossbars. We refer to the proposed technique as ReaLPrune. ReaLPrune is inspired by the recently proposed Lottery Ticket Pruning (LTP) hypothesis [11]. It combines the insights from LTP with the key attributes of the crossbar structure and the mapping strategy, guided by practical considerations adopted in ReRAMs for training CNNs. ReaLPrune can prune more than 90% of the CNN weights on average. The model pruned using ReaLPrune, can be trained from scratch with no accuracy loss using inexpensive hardware, compared to its unpruned counterpart. Moreover, due to its crossbar-aware nature, the resulting sparsity directly translates to a high amount of hardware (ReRAM crossbar) savings. ReaLPrune also outperforms existing pruning techniques (including crossbar- aware methods) in terms of achievable sparsity. This enables us to accelerate the training of deeper and larger CNNs on hardware constrained platforms (such as edge devices, i.e., edge AI). The key contributions of this work are as follows:

- We show that despite pruning more than 90% of the weights, LTP is unable to achieve similar levels of hardware savings or performance improvement in practice, for a ReRAM-based architecture.
- We propose a novel crossbar-aware pruning strategy, referred to as ReaLPrune. This strategy achieves more than 90% sparsity while reducing hardware requirements and enhancing performance significantly.
- Experimental analysis indicates that ReaLPrune-enabled training is ~20× faster than training with unpruned models on an ReRAM-based architecture.

The rest of the paper is organized as follows. Section 2 presents relevant prior work related to pruning and ReRAM- based architectures. Section 3 motivates the necessity of a crossbar-aware mapping strategy. Section 4 introduces the proposed ReaLPrune technique. We evaluate ReaLPrune's effectiveness in Section 5. Finally, we conclude this paper by summarizing the findings in Section 6.

## 2 RELATED PRIOR WORK

In this section, we present relevant prior work on ReRAM-based CNN accelerators and model pruning techniques.

### 2.1 ReRAM-based architectures

ReRAMs can be used to perform in situ multiply-and-accumulate (IMA) operation, which forms the core of CNN computational kernel. Hence, ReRAM-based architectures are popular for accelerating inferencing for CNNs [7]. A working prototype of ReRAM-based architecture for CNN inferencing has been demonstrated by researchers from CEA-Leti [8]. Recent work has attempted to design ReRAM-based systems for CNN training [6][10][14]. However, lower precision of computing, the lack of normalization layers, and endurance issues have presented a challenge towards adopting ReRAM based accelerators for CNN training [10][30][41]. It is well known that the weight gradients in a CNN are very sensitive to precision [10]. Hence, training with low precision representation can often lead to accuracy loss or failure to train altogether. The poor accuracy problem can be addressed by using a combination of ReRAMs and GPUs [15]. However, GPUs are relatively slower than ReRAMs for performing IMA operations. This can result in relatively sub-optimal performance. In [31], the authors solve the problem of training at low precision using stochastic rounding. In [10], the authors propose using GPUs to support normalization layers for training deep CNNs. However, all these ReRAM-based systems assume ideal ReRAM behavior. Due to immature fabrication process, ReRAM cells often have various types of faults [28][29]. Moreover, the frequent weight updates involved in CNN training can lead to new faults as ReRAM cells have relatively poor write endurance. CNN training and inferencing on non-ideal ReRAM crossbars can lead to accuracy drop [29][41]. Several methods including the use of error correction code (ECC), weight clipping, and selective weight updates, have been proposed to enable successful CNN training and inferencing even in presence of faulty ReRAM cells [29][42]. By incorporating these techniques in conventional ReRAM-based architectures, we can train CNNs with minimal accuracy drop even when many ReRAM cells are damaged/defective. We can adopt these measures in an ReRAM-based architecture for enabling reliable CNN training in the presence of faults and defects. However, all the above-mentioned architectures utilize conventional unpruned CNNs, which tend to have high area, relatively low performance, and high energy requirements. In this work, we demonstrate the potential of training already pruned CNNs on ReRAM-based architectures. Our experiments indicate that training pruned CNNs from scratch requires significantly fewer hardware resources and also reduces execution time, which meets the requirements of training on edge.

### 2.2 CNN Pruning

It is estimated that training a single unpruned neural network can cost over $10,000 and emit as much carbon as five cars over their lifetimes [19]. Pruning can solve this challenge by reducing the storage and energy requirements. It also accelerates both CNN training and inferencing. Several pruning techniques have been proposed in the literature [11][16][26][27][39]. However, all these techniques are unaware of ReRAM crossbar structure. The mapping of CNN weights to the ReRAM crossbars is very different than conventional GPUs. In a ReRAM crossbar, each input activates all the cells in a row of the ReRAM crossbar. Similarly, each output activates all the cells in a column of the ReRAM crossbar. Pruning techniques that are unaware of these features of the ReRAM crossbar, may not lead to any hardware savings or performance benefits as we explain and experimentally demonstrate in more detail later. Crossbar aware pruning techniques have been proposed recently [12][13][17][40]. However, all these methods are targeted for CNN inferencing and are not as effective for training. The networks pruned using these types of methods typically fail to reach the same accuracy as their unpruned counterparts, when trained from scratch [11]. Pruning methods for supporting faster CNN training have also been proposed [36][37]. However, these methods start with an unpruned CNN and then prune weights after each epoch/iteration of training. While this strategy can improve performance, it is not amenable to reducing the hardware requirements. Hardware design must be done considering the worst-case scenario. In this case, we need an ReRAM-based system that can support the unpruned CNN. Even though the CNN is pruned over the next few epochs/iterations, the hardware cannot be pruned/reduced at runtime, i.e., the additional cells will still remain in the design. Hence, there is no hardware savings following this strategy. In addition, the pruning has to be repeated every time the same CNN model needs to be trained from scratch, which is fundamentally different from what we aim to achieve. In this work, we want to prune before training. The pruned model can then be trained from scratch and/or incrementally.

Lottery Ticket Pruning (LTP) is a recently proposed pruning technique for CNN training that addresses these shortcomings in existing pruning methodologies [11]. The pruned model obtained using LTP can be trained from scratch with little to no accuracy loss when compared to the original unpruned model. Hence, we can use fewer ReRAM crossbars to train the CNNs. Moreover, the pruned CNNs are reusable any number of times, i.e., we can train the same pruned CNN over and over, thereby amortizing the cost of pruning itself. However, LTP method is oblivious of the ReRAM crossbar structure. Hence, despite pruning more than 90% of the CNN weights using LTP, we do not see commensurate hardware savings or performance gains in practice. In this work, we address the above shortcomings of existing ReRAM-based architectures and pruning techniques. We present ReaLPrune, an iterative crossbar-aware pruning technique, that removes weights strategically to save area, and improve performance and energy-efficiency. Experiments demonstrate that ReaLPrune achieves high sparsity (more than 90%) for a variety of CNNs while also leading to significant hardware savings and performance benefits.

## 3 LOTTERY TICKET PRUNING: CHALLENGES

In this section, we discuss LTP and explain why it is not effective for an ReRAM-based architecture.

### 3.1 Lottery Ticket Pruning (LTP)

LTP is a pruning technique that can find sparse sub-networks for CNN training [11]. LTP shows that dense, randomly initialized networks contain sparse subnetworks (referred as "winning tickets"), that when trained in isolation, reach test accuracy comparable to the original network using a similar number of iterations. LTP achieves significantly higher sparsity than existing pruning methods that uses regularization techniques such as L1/L2-norm or group lasso. Fig. 1 explains the difference between LTP and conventional pruning. Conventional pruning techniques (that are targeted for inferencing) first train an unpruned network, and then prune the unimportant weights. The pruned network is then used for inferencing with the pretrained weights. In some cases, the pruned network is further retrained (with weights initialized using the pretrained weights) to recover accuracy loss. However, it should be noted that these pruned networks typically fail to reach the same accuracy when trained from scratch [11]. Hence, conventional pruning techniques that are targeted for inferencing are often not effective for CNN training. Unlike conventional pruning techniques, LTP prunes the CNN before training as shown in Fig. 1. LTP first prunes the network; the pruned network can then be used for multiple training instances. CNN training is expensive in terms of both storage and computation. Training a pruned network from scratch has the potential to reduce computation and storage requirements as it involves fewer weights to store and train. This can significantly speed-up CNN training and is also more energy efficient.

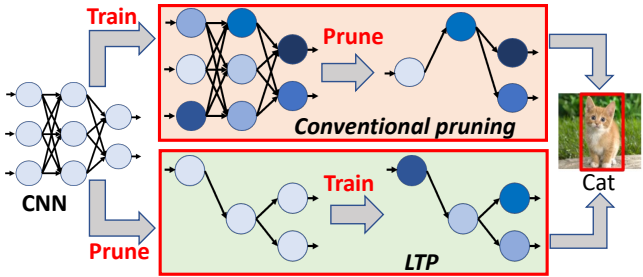

Fig. 1: Illustration of LTP and its difference with conventional pruning techniques.

In [36] and [37], the authors present two pruning techniques for training ML models. The PruneTrain method prunes the model during training from scratch using group lasso regularization. After every epoch, the weights are pruned, and the CNN is reconfigured to continue training on the pruned model. A similar strategy is adopted in [37]. The method proposed in [37] prunes weights after each epoch using a variable pruning threshold. In addition, this method incorporates the ability to recover incorrectly pruned weights in subsequent iterations. However, both these approaches prune the model during training. As mentioned earlier, it has the following drawbacks: (a) Unlike the CNN, the hardware elements cannot be physically removed depending on the CNN size. The hardware (ReRAM-based system in this case) must be designed considering the worst-case scenario i.e., the unpruned CNN. Hence, there is no hardware savings using this method, and (b) The process of pruning has to be repeated every time the CNN model is trained. This is fundamentally different from what we aim to achieve. In this work, we want to *prune the model before training*. The pruned model can then be trained any number of times from scratch. This has the potential to improve power and performance in addition to reducing hardware requirements, all of which are necessary to enable training on end user devices.

To identify a winning ticket, LTP adopts the following steps:
- Randomly initialize the neural network with parameters $\theta_0$ at time $t = 0$.
- Train the neural network for k iterations, resulting in parameters $\theta_k$.
- Prune p% of the smallest-magnitude weights. The parameter p can be chosen by the user.
- Reset the unpruned weights to its original initialization value (i.e., $\theta_0$). These remaining, unpruned weights constitute the winning ticket.
- Retrain the winning ticket using same data and repeat the above steps until MAX iterations

By repeating this process once (one-shot pruning) or in an iterative manner (iterative pruning), LTP can uncover winning tickets that are more than 90% sparse. These sparse models achieve high accuracy like their unpruned counterpart when trained from scratch. The iterative LTP consistently outperforms its one-shot counterpart for all CNNs [11]. The pruned sub-networks, also referred to as the "winning tickets", exhibit many interesting features:

- Aggressively pruned networks (with 95-99.5% of weights pruned) show no drop in accuracy while moderately pruned networks (50-90% pruning) often outperform their unpruned counterparts [11][21].
- The pruned networks meet/exceed the unpruned network's test accuracy within the same number of iterations [11].
- The winning tickets generalize across a variety of datasets (including Fashion MNIST, SVHN, CIFAR-10/100, ImageNet, and Places365), i.e., they are dataset agnostic [18].
- The lottery networks work equally well with different optimizers (such as SGD, Adam, etc.) with high accuracy [18].
- The lottery network can be easily trained using different hyper-parameter settings, especially if it is generated using larger datasets [18].
- Winning tickets can be identified at very early stages of training with aggressively low-cost training algorithms to reduce computation effort and runtime for LTP [20].
- We can stretch (or squeeze) the pruned network into another deeper (or shallower) network from the same family i.e., the pruned network characteristics are transferable across CNNs of the same family [38].

These features of LTP make it an attractive choice for pruning CNNs for the purpose of training. Hence, we choose the LTP strategy in this work for pruning. We can then train these pruned CNNs on ReRAM-based systems from scratch.

### 3.2 Challenges with LTP

Despite these advantages, LTP is not suited for ReRAM-based architectures as it is unaware of the crossbar structure and the mapping policy used to map CNN weights to the ReRAM cells. Fig. 2 explains this problem. As shown in Fig. 2(a), we consider a scenario where 12 out of the 16 ReRAM cells are zero (sparsity level: 75%). As every row/column in Fig. 2(a) has at least one non-zero entry, we cannot save any hardware. Here, we define 'hardware savings' as the fraction of ReRAM cells that can be simply turned off (by power gating) or reused for other purposes (e.g., mapping other non-zero weights of the CNN) without affecting the correctness of the intended MAC operation. However, a crossbar cannot control individual ReRAM cells to avoid storing the zero weights and save hardware. This happens as an input to the crossbar activates the entire row while the output is obtained by accumulating the sum of currents from all the cells along a single column. Therefore, if even one cell in a row/column has a non-zero value, any naïve attempt to turn off or reuse the other cells storing zeros, will produce incorrect results.

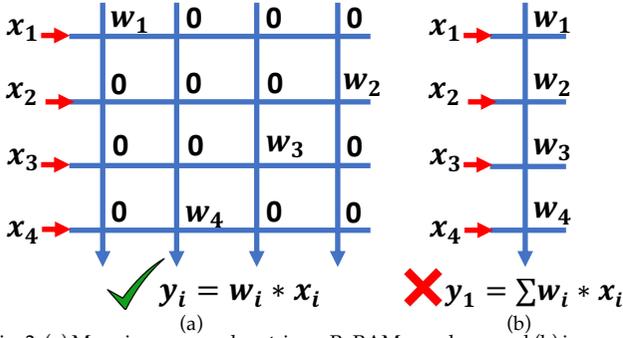

Fig. 2: (a) Mapping a pruned matrix on ReRAM crossbars, and (b) incorrect results obtained after rearranging the non-zero entries.

Similarly, compressing/rearranging the non-zero entries to save hardware is also ineffective. In Fig. 2(b), we have rearranged the remaining four non-zero entries (from Fig. 2(a)) on a single column to highlight the problem. As shown in Fig. 2(b), this results in an incorrect output (the correct and intended output $y_i$ is shown in Fig. 2(a)). Hence, we must preserve and use all the 16 cells as shown in Fig. 2(a) to ensure correct outputs. As a result, there is no hardware saving despite 75% sparsity in Fig. 2(a). We can only save ReRAM resource (cells) when an entire row/column is filled with zeros; such a row/column can be utilized for other purposes/computations without affecting the correctness of the intended output unlike Fig. 2(b). Hence, even though many cells in Fig. 2(a) are storing zeros, we must leave them as it is, i.e., there is no hardware saving. This observation indicates that crossbar-unaware sparsity does not proportionately translate to savings in hardware, especially for ReRAM-based systems.

This problem is exacerbated for larger crossbars. Typically, 128×128 crossbars are used for CNN training and inferencing [7][10][12]. Similar to Fig. 2(a), we cannot save any ReRAM resource if we have 128 non-zero entries (out of a total of 128×128=16384 entries, sparsity level: 99.2%), with each row/column having at least one non-zero value; note that this represents a worst-case scenario. Overall, crossbar-unaware pruning strategies, such as LTP, are not suited for ReRAM crossbars as they may not lead to significant hardware savings despite high amount of pruning. Hence, a suitable pruning strategy for ReRAM-based systems must be aware of the crossbar structure and the mapping strategy adopted to represent weights on ReRAM cells.

## 4 CROSSBAR-AWARE REALPRUNE

In this section, we first discuss the important features of the ReRAM crossbars that govern the formulation of the ReaLPrune technique. Next, we present the overall training process that incorporates ReaLPrune for crossbar-aware LTP.

### 4.1 Crossbar awareness

A typical Conv layer operation in a CNN has a total of *OC* filters, where each filter is of shape *IC×K×K*. The parameters IC and OC represent the number of channels in the input and output of the convolution layer, respectively. The input to a Conv layer is a tensor of shape *IC×I×I* while the output is of shape *OC×O×O*. The output is obtained after multiplying the weights with the inputs. The parameters *I* and *O* represent the dimensionality of the input and output of a convolution layer respectively. Fig. 3(a) shows how the weights of a Conv layer are mapped to ReRAM crossbars. Some Conv layers have millions of trainable weights, which cannot be mapped to one ReRAM crossbar; each crossbar typically stores a maximum of 128×128 entries. Hence, the weights of a Conv layer are mapped on to multiple ReRAM crossbars as shown in Fig. 3(a). From Fig. 3(a), we note that to save a column in an ReRAM crossbar (i.e., all entries in the same column are zero), we must prune one (or more) channels of a filter (Channel-wise pruning as shown in Fig. 3(c)). Pruning an entire filter (of shape *IC×K×K*) also achieves similar results (Filter-wise pruning as shown in Fig. 3(b)). As mentioned earlier, 'saving a row/column' implies that all the ReRAM cells in the saved row/column can be freely reused for other purposes without affecting the output of the MAC operation as shown in Fig. 2. Similarly, to save a row in a ReRAM crossbar (i.e., all entries in a row are zero), we must prune multiple (or all) weights at the same index for all the filters (index-wise pruning as shown in Fig. 3(d)). We use these insights to develop ReaLPrune.

Next, unlike inferencing, CNN training involves an additional backward phase for calculating error/weight gradients. The error/weight gradient calculations require storing the activations from the forward phase. Therefore, ReaLPrune must also prune the activations to reduce the total

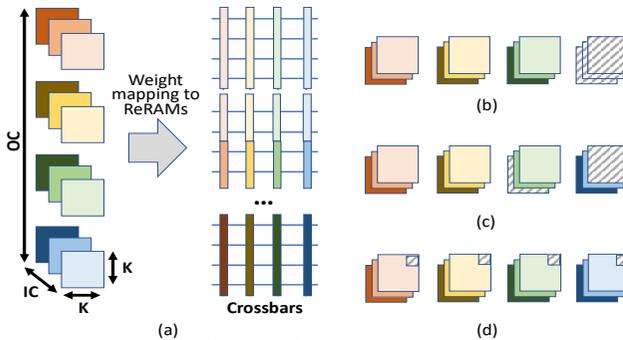

Fig. 3: Illustration of (a) how weights of a Conv layer are mapped on to ReRAM crossbars, and (b) Filter-wise, (c) Channel-wise, and (d) index-wise pruning strategies adopted in ReaLPrune. (Gray shaded regions indicate which regions are pruned in each of the strategies).

**Algorithm 1.** Pruning using ReaLPrune

**Input**: CNN model, pruning percentage $p$
**Output**: Pruned CNN model
**Algorithm**:

| | |
|---|---|
| 1: | **Initialize**: $w \leftarrow w_{initial}$; |
| 2: | **While** $itr$ < MAX_ITER and no accuracy drop**:** |
| 3: |    **Train** for $E$ epochs |
| 4: |    **Prune** ($p$) based on the crossbar structure and magnitude of weights |
| 5: |    **If** *New_accuracy* < *Baseline_accuracy* |
| 6: |       Undo last pruning step |
| 7: |       Switch to finer pruning strategy |
| 8: |    **Reinitialize** remaining weights with $w_{initial}$ |
| 9: | **Return** *Pruned Model* |

ReRAM requirements. However, pruning activations using LTP (which prunes only the weights) is not straightforward. Activations are input-dependent and can be pruned only when the entire filter is filled with zeros; pruning an entire weight filter (of shape $IC{\times}K{\times}K$) causes an output channel to vanish (i.e., pruned). On the other hand, the pruning of only one (or few) channels/indices of the weights does not result in a zero activation. Therefore, ReaLPrune must prioritize pruning entire filters to simultaneously reduce the number of weights and activations that must be stored on ReRAMs.

Here, it should be noted that the crossbar-aware pruning (as shown in Fig. 3) is different from traditional structured pruning (such as [27]) that are targeted for GPU-based platforms. For instance, the structured pruning method in [21] prunes a channel from all the filters of a CNN layer. However, as shown in Fig. 3(c), that is not the case for crossbar-aware pruning. As shown in Fig. 3(c), pruning a channel in one filter does not necessitate pruning the corresponding channel of all the other filters to reduce hardware requirements. Similarly, in the case of index-wise pruning, we do not need to prune the same index of all the filters at the same time.

### 4.2 ReaLPrune technique

Similar to LTP (Fig. 1), ReaLPrune has two stages: (a) Pruning the neural network, and (b) In-field training (deployment) of the pruned model. In this sub-section, we present the pruning phase of ReaLPrune. As shown in Fig. 3(b)-(d), ReaLPrune prunes (a) filter-wise, to reduce the number of ReRAM cells required for storing both the activations and weights, (b) channel-wise, to ensure that one or more columns in a ReRAM crossbar is filled with zero, and (c) index-wise, to prune all entries along the same row in a crossbar. To maximize the amount of pruning, ReaLPrune adopts a coarse-to-fine pruning strategy, i.e., we start by pruning filter-wise (the coarsest granularity of pruning), followed by channel-wise and then finally index-wise pruning (the finest granularity of pruning). We prioritize and initiate ReaLPrune with filter-wise pruning as it is the only pruning strategy that reduces both activations and weights. However, due to its coarse granularity, it does not lead to significant amount of pruning without sacrificing accuracy. Hence, we gradually shift towards finer granularity of pruning to ensure maximum possible sparsity for achieving the same accuracy as the unpruned variant. We present the high-level details of ReaLPrune in Algorithm 1.

Algorithm 1 shows the overall training process using ReaLPrune. As shown in Algorithm 1, the input to ReaLPrune is the CNN model and the percentage of weights (p) we want to prune after each iteration. The output of ReaLPrune is the pruned CNN model, which when trained in isolation from scratch, will lead to comparable accuracy as its unpruned counterpart. We begin by initializing the CNN model ($w_{initial}$ at $t = 0$) as shown in Line 1 of Algorithm 1. We can use any of the commonly used initialization schemes here e.g., Xavier, Kaiming, etc. Next, we perform the following steps: (a) train the model for $E$ epochs (Line 3), (b) Prune the lowest $p$ percentile of non-zero weights by magnitude following the crossbar-aware coarse to fine pruning strategies (i.e., filter-wise, channel-wise and index-wise) (Line 4), (c) if the testing accuracy of pruned model is lower than the baseline accuracy (for unpruned CNN), undo last pruning and shift to finer pruning strategy (Lines 5-7); if the accuracy drop is zero, then no action is necessary, (d) Reinitialize the network (Line 8 of Algorithm 1) with $w_{initial}$ from $t = 0$ (except the pruned weights) and repeat Steps (a)-(d). The pruning repeats until MAX iterations are reached or until there is an accuracy drop. The CNN model that has the maximum amount of sparsity with no accuracy drop is returned by the algorithm.

The ReaLPrune technique follows the iterative magnitude pruning with the reinitialization strategy, to reach the sparse model; this strategy is inspired by LTP. As mentioned earlier, LTP also adopts an iterative magnitude pruning strategy where the lowest $p$ percentile of weights (by magnitude) is pruned after each training iteration. However, LTP prunes weights without considering their locations. In contrast, ReaLPrune is crossbar-aware. As an example, we prune an entire filter (in filter-wise pruning) if the average weight of that filter is among the lowest $p$ percentile considering all the filters of the CNN. Similarly, we prune an entire channel (in channel-wise pruning) or the same index on multiple filters (in filter-wise pruning) if the average weight of that channel or that index, is among the lowest $p$ percentile respectively. By repeatedly pruning a small fraction (lowest $p$ percentile) of the weights in each iteration, following the coarse-to-fine strategy, ReaLPrune is able to prune more than 90% of the weights. These savings directly translate to hardware savings and better performance.

### 4.3 Mapping ReaLPrune to Hardware

In this sub-section, we discuss the implementation of ReaLPrune (including its in-field deployment) on hardware. As shown in Algorithm 1, the pruning step requires iterative training (Line 3 of Algorithm 1). Here, we first emphasize that this training (for obtaining the pruned model)

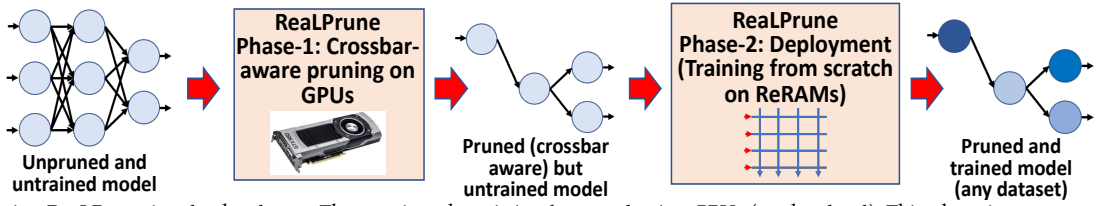

Fig. 4: Deploying ReaLPrune in edge hardware. The pruning phase is implemented using GPUs (on the cloud). This phase incorporates the crossbar knowledge (i.e., which weights will be mapped to which crossbar) for pruning; the deployment phase (in-field training) is done using ReRAM crossbars

is separate from the in-field training during the deployment phase. The pruning phase of ReaLPrune is a one-time process only. Hence, the training required for crossbar-aware pruning (Line 3 of Algorithm 1) can be implemented using other hardware alternatives (e.g., GPUs, TPUs, etc.); this step is not being implemented on ReRAMs. Once, the pruned model is obtained, the remaining weights are rewound to their original values at $t = 0$, and then deployed on ReRAM crossbars for all future training instances with any dataset. Recall that models pruned using LTP can be trained from scratch without any accuracy loss. This enables us to train deep and large CNNs on hardware constrained platforms (such as edge devices).

Fig. 4 illustrates the implementation of the different stages of ReaLPrune on suitable hardware platforms. As shown in Fig. 4, the pruning (and the associated training; Line 3 of Algorithm 1) is implemented using conventional GPUs. This step incorporates the crossbar knowledge (i.e., which weights are mapped to which crossbar) during the pruning phase as discussed earlier in Sec. 4.1. Note that the mapping of CNN weights to ReRAM crossbars is deterministic following [6]. For instance, all weights belonging to the same filter in a Conv layer are mapped to the same crossbar column [6]. We incorporate such mapping information in phase 1 (pruning step) of ReaLPrune. Once, the pruned model is obtained, we map the remaining weights to ReRAM crossbars as shown in Fig. 5. Fig. 5(a) shows the target ReRAM-based on-chip training hardware; we discuss the architecture is more detail in next section. Fig. 5(b) shows an example where 16 weights are mapped to a 4×4 ReRAM crossbar. Four weights ($w_3$, $w_7$, $w_{11}$, $w_{15}$), all belonging to the same column, are pruned (i.e., denoted by the red color). As a result, we can reuse these four cells for other computations (e.g., map another set of weights), without affecting accuracy. This results in significant hardware savings and performance improvement as we show later. Overall, the ReRAM crossbars are used for training the pruned model in-field, from scratch (i.e., deployment phase of ReaLPrune); the pruning itself need not be implemented using ReRAMs.

## 5 EXPERIMENTAL RESULTS

In this section, we first compare ReaLPrune with other pruning techniques in terms of network sparsity and hardware savings. Next, we present results on the full-system speed-up for in-field training enabled by ReaLPrune.

### 5.1 Experimental setup

The pruning phase of ReaLPrune is implemented using NVIDIA Titan Xp GPU with 24GB of memory. The pruned (but untrained) network is then mapped to a manycore ReRAM-based PIM architecture for evaluating in-field training speed-up and hardware savings. Fig. 5 shows the target hardware platform, which consists of multiple ReRAM tiles. Each ReRAM tile can be configured for both storage and computation. Each tile includes eDRAM buffers, IMA units, output registers, along with shift-and-add, ReLu, and max-pool units. The IMAs have multiple crossbar arrays along with other peripheral circuitry, e.g., ADCs, connected with a shared bus. In line with prior work [7][10], 16-bit fixed-point precision is used for the computations on ReRAMs. The specific embodiment of the target architecture considered in this work consists of 256 ReRAM tiles. The tiles are connected using a mesh network-on-chip (NoC). Here it should be noted that mesh NoCs are not typically suited for multi-hop long-range communication. However, CNN training involves data sharing between adjacent layers only. Hence, long-range communication can be avoided by appropriately mapping the CNN layers to different processing tiles [10]. As a result, a simple NoC topology such as mesh is sufficient as the communication backbone in ReRAM-based architectures. Each ReRAM tile consists of 96 crossbars (each crossbar is of size 128×128) and the associated peripherals such as ADC, DAC, etc. Each ReRAM tile requires 0.37 mm² area and consumes 0.33 W power [7]. The ReRAM crossbars operate at 10 MHz. We use NeuroSim V2.0 to evaluate full-system area and performance of the ReRAM architecture after mapping the pruned CNN model [31]. NeuroSim V2.0 provides support for on-chip training and includes hardware for feed-forward, error-calculation, weight-gradient-calculation and weight-update. Hence, it is suitable for evaluating CNN training on ReRAM-based systems.

**CNNs used for evaluation**: We choose four well known CNNs: VGG-11, VGG-16, VGG-19, and ResNet-18 for experimental analysis [23][24]. The CNNs are trained on the CIFAR-

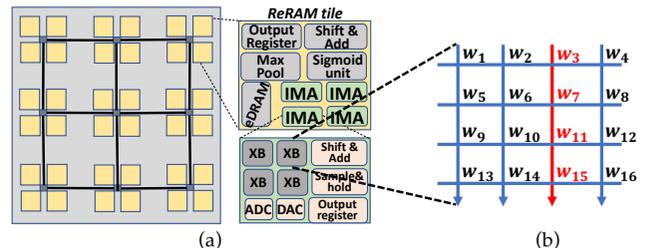

Fig. 5: Illustration of the (a) ReRAM-based architecture for CNN training, and (b) mapping the pruned weights on the ReRAM crossbar

10 dataset as an example [25]. However, as discussed earlier, the pruned lottery network generalizes across a variety of datasets (including Fashion MNIST, SVHN, CIFAR-10/100, ImageNet, and Places365) [18]. We show in Sec. 5.5 that similar to LTP, ReaLPrune is highly effective for different datasets and other deeper CNNs as well. The CNNs were implemented using PyTorch and trained on a NVIDIA Titan Xp GPU with 24GB of memory. The training was done using Xavier initialization, learning rate (LR) of 0.1, batch size of 128, and SGD optimizer. The LR was decreased by 5% after every epoch. The use of Xavier initialization enables us to successfully train the deep CNNs. We train all the CNNs for 50 epochs. Here, it should be noted that both the unpruned and pruned CNNs take similar time to reach same accuracy [11].

**Baseline pruning techniques**: As discussed in Section 2, there are multiple ways to prune a CNN. We can prune during training as done in [36][37]. However, as mentioned earlier, the objective of our work is to prune before in-field training (as shown in Fig. 4); hence, these methods are not suitable as baselines here. A reinforcement learning (RL) based pruning approach is proposed in [40]. However, this approach is not suitable for CNN training as the RL model has to be retrained every time the weights change (due to weight updates). The most common method of pruning requires adding a regularization term to the loss function, such as the L1-norm of weights or a group LASSO that uses L1-norm or L2-norm of groups of weights for structured pruning [16][17][27]. The regularizer penalizes complex models and prefers simpler models which perform well on the training data and leads to better generalization in both theory and practice. This causes the optimization process to automatically prefer small absolute values for weights or groups of weights. The less important weights become zero (or too small) in the process, thereby sparsifying the model. We choose two different pruning techniques from this family of pruning methods as representatives, to compare with ReaLPrune. We choose a block pruning technique (referred as 'Block' hereafter) that uses group LASSO to prune blocks of weights [16]. We adapt this technique for the crossbar configuration in our target architecture. Also, we employ a recently proposed crossbar-aware pruning (referred as 'CAP') technique [13]. CAP utilizes a multi-group LASSO algorithm to prune groups of weights that would otherwise be mapped along a column in an ReRAM crossbar. Here, it should be noted that these two pruning approaches achieve similar levels of pruning as the other methods (such as [12][17][27]); hence, they are suitable as baselines to evaluate the effectiveness of ReaLPrune. Both Block and CAP are implemented in an iterative manner to ensure maximum possible pruning without sacrificing accuracy compared to their unpruned counterparts. We also choose LTP as the representative state-of-the-art crossbar-unaware pruning technique as the third baseline as it achieves one of the highest levels of sparsity among the pruning techniques considered here [11]. We prune 25% of the remaining non-zero weights after each iteration based on their magnitude (i.e., $p = 0.25$ in Algorithm-1). Please note that the pruning percentile ($p$) is a hyperparameter (similar to learning rate), and the value of $p$ can be decided by the user.

**Reliability of training**: In this work, we assume ideal ReRAM behavior. However, as mentioned earlier, non-ideal ReRAMs have many shortcomings that can affect the quality of CNN training, such as the use of low precision (16-bit fixed point in this case) and write endurance. These issues can be addressed using very simple techniques. To mitigate the accuracy loss due to the use of 16-bit fixed point precision, stochastic rounding can be used [10]. Stochastic rounding is an unbiased rounding scheme that makes a probabilistic decision of where to round and has the desirable property that the expected rounding error is zero. The use of stochastic rounding leads to successful CNN training at less than 1% area overhead [10]. To address the problem of write endurance, we can adopt the low rank training (LRT) algorithm proposed in [9], which reduces the number of weight updates by two orders of magnitude. As an example, training a CNN for 50 epochs on CIFAR-10, with batch size of 128, results in ~20k weight updates. Prior work has reported ReRAM write endurance between $10^6$-$10^{12}$ writes [43][44]. Even if we assume the most pessimistic scenario of $10^6$ writes, the LRT method will allow us to train more than 5000 times assuming that the training configuration (e.g., batch size, number of epochs, and dataset size) remains the same. Alternatively, we can adopt a magnitude-based weight update method as outlined in [42]. Either of these methods can be used to reduce the number of writes necessary for CNN training. In spite of all these measures, faults can still happen due to a variety of reasons [28][29]. We can adopt additional counter measures such as ECC, and weight clipping to continue reliable training. For instance, the use of weight clipping enables successful CNN training with up to 5% fault density, while introducing significantly low overheads [29]. Hence, we can train the pruned CNNs obtained using ReaLPrune, even if the ReRAMs were non-ideal. However, since addressing reliability issues in ReRAMs is not the focus of this work, we will assume ideal behavior for demonstrating the effectiveness of ReaLPrune and for all further analysis.

### 5.2 Accuracy after pruning

First, we compare the effectiveness of each of the pruning techniques in terms of the achievable sparsity. For this purpose, we first prune the CNNs using the four different methods to obtain the respective sparse networks. Here, our goal is to find the sparsest possible CNN that can be trained from scratch to achieve prediction accuracy that is at par or higher than the baseline accuracy. We define 'baseline accuracy' as the accuracy obtained after training the original unpruned CNN model. Fig. 6(a) shows the amount of pruning that each technique can achieve without sacrificing accuracy for all the four CNNs. As shown in Fig. 6(a), all these techniques can prune a significant percentage of weights when they are applied in an iterative fashion. LTP performs the best and can prune 97.2% of the weights on average. ReaLPrune can prune 95.5% of the weights on average for the four CNNs considered here. Block and CAP prune 87.3% and 87.5% of weights on average, respectively. Models with higher levels of pruning (than what we report here) failed to reach baseline

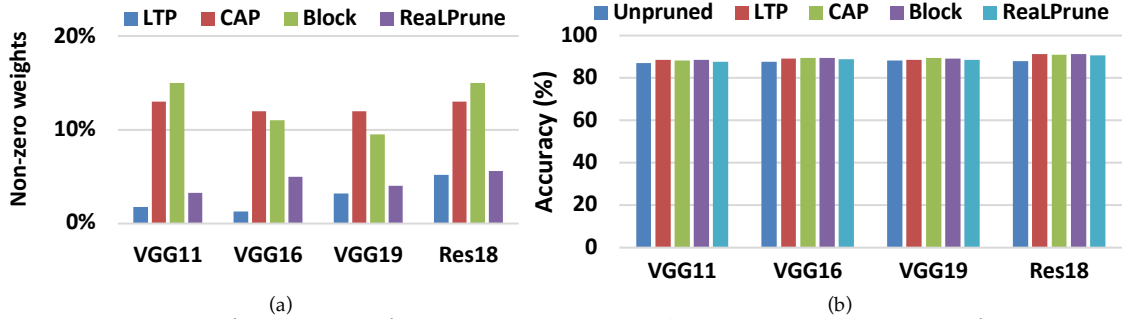

Fig. 6: (a) Percentage of non-zero weights remaining in the models after pruning, and (b) Accuracy obtained after training the pruned models for 50 epochs from scratch simulating an on-field training

accuracy when trained from scratch. ReaLPrune can remove more weights than Block and CAP as it adopts multiple pruning strategies (filter-wise, channel-wise, and index-wise pruning) as discussed in Sec. 4.2; both Block and CAP adopt a singular pruning strategy (row/column-wise pruning) [6][10]. As shown in Fig. 6(a), LTP is the clear winner in terms of the achievable sparsity for all CNNs. However, as we show in next sub-section, it fails to reduce hardware requirements proportionate with the high levels of pruning.

Once we have the pruned models, we can then use them for in-field, on-chip training. Fig. 6(b) shows the accuracy achieved by the sparse models (from Fig 6(a)) compared to the baseline accuracy (i.e., accuracy of the unpruned model) when trained from scratch for 50 epochs simulating an in-field, on-chip training. As shown in Fig. 6(b), all the pruned models achieve at par or slightly higher accuracy than their unpruned counterpart when trained for 50 epochs. For instance, the unpruned ResNet-18 model achieves 87.95% accuracy when trained for 50 epochs. Following the same training configuration, the ReaLPrune-enabled model achieves 90.66% accuracy even when 94.4% of its weights are pruned. This happens as pruning acts as a regularizer, which makes the sparse model generalize better on unseen data (as shown in Fig. 6(b)). However, we see a steep decline when we continue to prune more weights than what we report in Fig. 6(a). This happens as the network now has very few parameters and is unable to learn all meaningful representations from the input. These observations are also in line with prior work [11]. This experiment shows that we can use these suitably pruned models for on-chip training on the edge without accuracy loss.

## 5.3 Hardware savings due to pruning

As shown in Fig. 4, crossbar rows/columns with all zero weights can be reused for other purposes without affecting the intended output i.e., we will need fewer ReRAM cells to train the pruned model compared to the unpruned baseline. Fig. 7 shows the number of ReRAM crossbars that are necessary to train the pruned models (from Fig. 6(a)) compared to their unpruned counterparts. Here, it should be noted that the number of ReRAM crossbars can vary based on the amount of parallelism adopted for accelerating the CNN training [6]. In a pipelined training implementation, the slower CNN layers will dominate the execution time. Hence, these slower layers are typically accelerated by replicating the weights using additional ReRAM crossbars [6]. To ensure fair comparison in terms of the hardware savings, we choose an iso-performance setting, i.e., we ensure equal amount of parallelism, and hence equal performance, for the four pruning techniques.

As shown in Fig. 7, under an iso-performance setting, the number of required ReRAM crossbars is significantly reduced by all the four pruning techniques. However, ReaLPrune achieves the highest amount of hardware savings despite pruning fewer weights than LTP. From Fig. 7, we note that ReaLPrune reduces the number of ReRAM crossbars necessary for training by 77.2% on average. LTP reduces hardware requirements by only 58.9% on average due to its crossbar-unaware nature, despite pruning more weights than ReaLPrune (Fig. 7(a)). Interestingly, Block and CAP achieve similar levels of hardware savings as LTP, despite pruning significantly fewer weights. Block and CAP reduces hardware requirements by 58.7% and 59% respectively. This happens as Block and CAP are crossbar-aware; hence, they can reduce hardware requirements despite pruning fewer weights.

We note that the amount of hardware savings is always less than the amount of pruning. This is expected as not all the pruned weights lead to hardware savings, as demonstrated in Fig. 2. In addition, ReRAM crossbars are required to store both the CNN weights and activations. However, weights and activations are pruned by different extents. This happens as only filter-wise pruning can prune activations (as shown in Fig. 3); channel-wise, index-wise, or other pruning strategies do not lead to an activation being zero. Hence, fewer activations are pruned than weights. As a result, the overall amount of hardware savings is always less than the amount of pruning despite the crossbar-awareness.

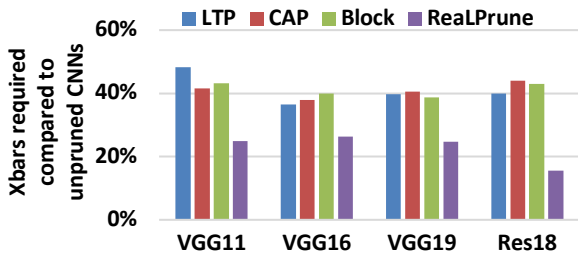

Fig. 7: Number of crossbars required for on-chip training after pruning

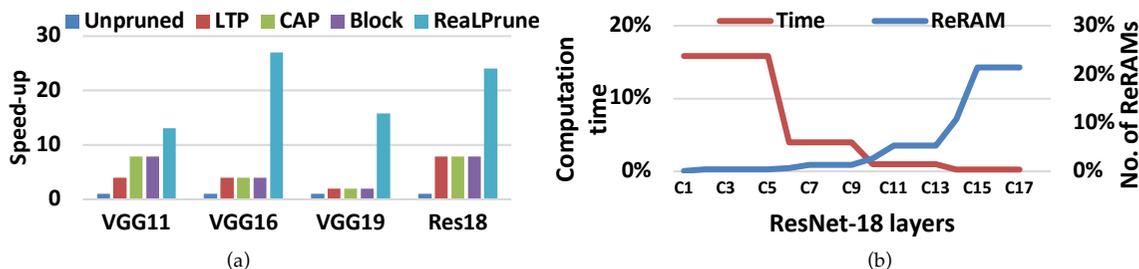

Fig. 8: (a) Overall execution time speed-up achieved by the different pruned models compared to unpruned CNNs, when training is done using ReRAMs, and (b) Layer-wise ReRAM crossbars requirement and corresponding execution time for ResNet-18.

## 5.4 Impact on performance

Next, we compare the speed-up enabled by the different pruning techniques compared to their unpruned counterpart on the same ReRAM-based architecture. Here, we do not show any performance comparison with respect to GPUs because prior work [6][14][33] has already demonstrated that ReRAMs are more efficient than GPUs for CNN training. Unlike the experiment in Section 5.2, we assume an iso-area setting for this analysis i.e., the number of ReRAM crossbars available is equal in all cases. Fig. 8(a) shows the speed-up when the different CNNs are trained using ReRAMs. As shown in Fig. 8(a), all the pruned CNN models lead to faster training than the unpruned model. ReaLPrune outperforms the other methods and achieves the highest speed-up (19.7×) for CNN training on average (under iso-area conditions) compared to the unpruned version. Fig. 8(b) explains this observation in more detail. It shows the minimum number of ReRAM crossbars and the corresponding computation time required by each convolution layer of an unpruned ResNet-18 (denoted as C1-C17). The CNN layers C11-C17 use up more than 80% of the ReRAM crossbars to store their weights, leaving only a handful of crossbars for the other layers. However, these layers (C11-C17) process smaller sized inputs than C1-C5. As shown in Fig. 8, the computation time for the first few layers (C1-C5) are the highest even though the number of weights associated with these layers are limited.

It is well known that CNNs are trained following a pipelined implementation on ReRAM-based architectures [6]. In a pipelined implementation, the slowest layer determines the overall execution time; in the case of unpruned ResNet-18, layers C1-C5 dominate the execution time. Hence, the slower layers must be accelerated by replicating the weights on additional ReRAM crossbars. However, in an unpruned ResNet-18, very few crossbars are available for these computation-heavy layers (more than 80% of the resources are used up for storing the weights of the layers C11-C17); this leads to lower speed-up in training. Unlike the unpruned model, ReaLPrune (and other pruning techniques) reduces the number of weights that need to be stored. This leads to significant hardware savings compared to the unpruned variant as shown in Fig. 7. In an iso-area setting, hardware savings translate to unutilized ReRAM crossbars. By using these available resources, we can accelerate the slower CNN layers. This leads to significantly higher speed-up for ReaLPrune, despite using the same total number of ReRAM crossbars as its unpruned counterpart (iso-area).

Finally, it should be noted that pruning (using either ReaLPrune, or the other techniques considered in this work) is a one-time effort. As shown in Fig. 1 and Fig. 4, our aim is to prune first, and then use the pruned model for all future training. We perform the iterative pruning using ReaLPrune only once; this step can be implemented using GPUs, TPUs, etc. The pruned model can then be made available publicly (e.g., via GitHub) for anyone to download and use. These pruned models can then be deployed on ReRAMs and reused for training (and inferencing) any number of times, thereby amortizing the cost (time/energy spent) for the pruning itself.

## 5.5 Scalability of ReaLPrune

In this sub-section, we show that ReaLPrune is equally effective for larger datasets and CNNs. For this experiment, we choose three different datasets, namely SVHN, CIFAR-100, and Tiny ImageNet [25][34][35]. SVHN includes 73257 images of digits for training, 26032 images of digits for testing, and includes 10 classes. The dataset represents a significantly harder, unsolved, real-world problem (recognizing digits and numbers in natural scene images) and is obtained from house numbers in Google Street View images [34]. The CIFAR-100 dataset has 100 classes containing 600 images each. There are 500 training images and 100 testing images per class [25]. Tiny ImageNet is a subset of the ImageNet dataset from the well-known ImageNet Large Scale Visual Recognition Challenge (ILSVRC). The dataset contains 100,000 images of 200 classes (500 for each class) downsized to 64×64-colored images. Each class has 500 training images, and 50 test images [35]. It should be noted that other datasets (such as ImageNet) can also be used here. However, training repeatedly on the entire ImageNet dataset from scratch is prohibitively expensive. Hence, we refrain from using ImageNet in this work noting that our experiments on diverse small and large datasets provide strong demonstration of our key research hypotheses. The datasets chosen here, have varying number of classes (SVHN: 10 classes, CIFAR-100: 100 classes, and TinyImageNet: 200 classes). This is necessary to demonstrate the scalability of LTP and ReaLPrune with increasing task complexity. For training with these three datasets, we use ResNet-18 as the underlying CNN. We only modify the final fully connected layer in ResNet-18 to account for the different image size and the different number of classes in each dataset. Here, we choose ResNet-18 as an example only. Similar

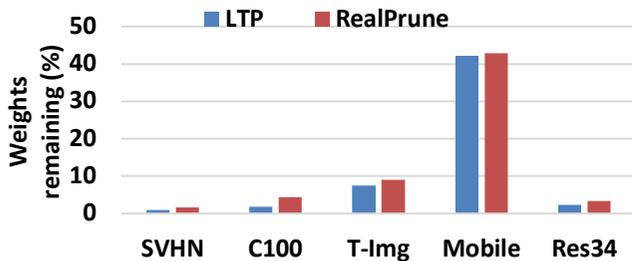

Fig. 9: Fraction of weights remaining after pruning using ReaLPrune compared to LTP for larger datasets, and deep CNN models. (C100: Cifar-100, T-Img: TinyImageNet, Mobile: MobileNet, Res34: ResNet-34)

observations are made with other CNNs as well.

In addition to new datasets, we also choose two relatively deeper CNN models, namely MobileNet [45], and ResNet-34. MobileNet is a simple and not very computationally intensive CNN with 28 layers, targeted for mobile vision applications. The MobileNet architecture factorizes a standard convolution into a depthwise convolution, and a 1 × 1 convolution called a pointwise convolution to reduce the model size. MobileNet is significantly smaller in size than VGGs despite having more layers. Hence, it is more challenging to prune. ResNet-34 is the deepest CNN considered in this work with 34 layers. Both these models are used with the CIFAR-10 dataset for this evaluation. By studying the behavior of ReaLPrune on these two larger models, we demonstrate the scalability of the proposed method for deeper CNNs.

From previous experiments (Fig. 6), we have demonstrated that both Block and CAP methods do not achieve comparable levels of sparsity as LTP and ReaLPrune. Hence, for this analysis, we only focus on LTP and ReaLPrune. Fig. 9 shows the percentage of weights remaining after pruning using LTP and ReaLPrune for the different datasets and deep CNNs. As shown in Fig. 9, the amount of sparsity that can be achieved after pruning decreases as the complexity of dataset increases. For instance, LTP was able to prune ~99% of the weights for SVHN (simplest of the three datasets) while it can only achieve 92.5% sparsity for TinyImageNet (most complex among the three datasets). This is expected as more weights are necessary to extract/learn distinguishing features among the different images of the more complex datasets. Interestingly, ReaLPrune achieves LTP-like sparsity for all the datasets irrespective of its size/complexity. As expected, LTP prunes slightly higher number of weights than ReaLPrune (which is similar to the observations in Fig. 6). Similarly, ReaLPrune is equally effective on deeper CNNs such as MobileNet (28 layers) and Resnet-34 (34 layers). In both cases, ReaLPrune achieves similar amount of pruning as LTP. Here, it should be noted that both LTP and ReaLPrune are not able to prune a lot of weights in MobileNet as shown in Fig. 9. This happens as MobileNet, by design, has fewer parameters to begin with. Hence, it is challenging to achieve extreme sparsity in MobileNet similar to the other CNNs. Overall, Fig. 9 shows that ReaLPrune is equally effective as a crossbar-aware pruning technique, irrespective of the dataset and CNN size for enabling high-performance on-chip training.

## 6 Conclusions

CNN training is expensive in terms of both the storage and computation requirements. Training a pruned network (from scratch) can alleviate this problem. However, existing crossbar unaware pruning techniques are not suited for this purpose. To address this problem, we have described a crossbar-aware pruning technique called ReaLPrune that achieves extreme sparsity (comparable to lottery ticket pruning), while also providing considerable savings in hardware. Our analysis has shown that ReaLPrune can prune 95.5% of CNN weights on average, which reduces hardware requirements by 77.2% compared to the unpruned version on average. In addition, ReaLPrune achieves 19.7× speed-up in execution time compared to the unpruned version on an ReRAM-based manycore architecture. ReaLPrune also outperforms other state-of-the-art pruning techniques, including crossbar-aware ones, in terms of both execution time and hardware savings.

## Acknowledgment

This work was supported in part by the US National Science Foundation (NSF) under grants CNS-1955353, CNS-1955196, and by the USA Army Research Office grant W911NF-17-1-0485. Biresh Kumar Joardar was also supported in part by NSF Grant # 2030859 to the Computing Research Association for the CIFellows Project.

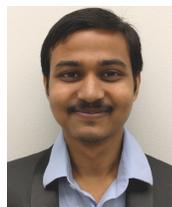

**Biresh Kumar Joardar** (M'20) finished his PhD from Washington State University in 2020. He is currently a Post-doctoral Computing Innovation Fellow (CI-Fellow) at the Department of Electrical and Computer Engineering at Duke University. His current research interests include machine learning, manycore architectures, accelerators for deep learning, hardware reliability and security. He received the 'Outstanding Graduate Student Researcher Award' at Washington state University in 2019. His works have been nominated for Best Paper Awards at prestigious conferences such as DATE and NOCS. He is a member of the IEEE.

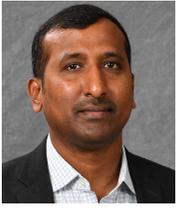
**Janardhan Rao Doppa (**M'14) is currently a George and Joan Berry Chair Associate Professor with Washington State University (WSU), Pullman, WA, USA. His current research interests are at the intersection of machine learning and computing systems design. He received a NSF CAREER Award (2019), an Outstanding Paper Award at the AAAI (2013) conference, a Google Faculty Research Award (2015), the Outstanding Innovation in Technology Award from Oregon State University (2015). He received the Reid-Miller Teaching Excellence Award (2018) and the Outstanding Junior Faculty in Research Award (2020) from the Voiland College of Engineering and Architecture at WSU. He is among the 15 outstanding young researchers selected to give Early Career Spotlight talk at the International Joint Conference on Artificial Intelligence (2021).

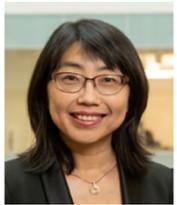
**Hai (Helen) Li** (M'08-SM'16-F'19) received her bachelor's and master's degrees from Tsinghua University, China, and her Ph.D. degree from Purdue University, USA. She is Clare Boothe Luce Professor and Associate Chair of the Electrical and Computer Engineering Department at Duke University. Her research interests include neuromorphic computing systems, machine learning and deep neural networks, memory design and architecture, and cross-layer optimization for low power and high performance. She has authored or co-authored more than 250 technical papers in peer- reviewed journals and conferences and a book. She received 9 best paper awards from international conferences. Dr. Li serves/served as an Associate Editor of a number of IEEE/ACM journals. She was the General Chair or Technical Program Chair of multiple IEEE/ACM conferences. Dr. Li is a Distinguished Lecturer of the IEEE CAS society (2018-2019) and a distinguished speaker of ACM (2017-2020). Dr. Li is a recipient of the NSF Career Award, DARPA Young Faculty Award (YFA), TUM-IAS Hans Fischer Fellowship from Germany, and ELATE Fellowship (2020). Dr. Li is an IEEE fellow and a distinguished member of the ACM.

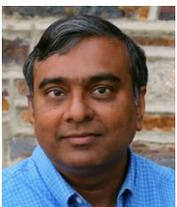
**Krishnendu Chakrabarty** received the Ph.D. degrees from the University of Michigan, Ann Arbor, in 1995. He is now the John Cocke Distinguished Professor and Department Chair of Electrical and Computer Engineering (ECE) at Duke University. His current research projects include: design-for-testability of integrated circuits and systems (especially 3D integration and system-on-chip); AI accelerators; microfluidic biochips; hardware security; machine learning for healthcare; neuromorphic computing systems. He is a Fellow of ACM, IEEE, and AAAS, and a Golden Core Member of the IEEE Computer Society.

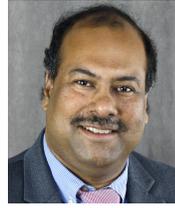
**Partha Pratim Pande** (M'05-SM'11-F'20) is a Professor and holder of the Boeing Centennial Chair in computer engineering at the school of Electrical Engineering and Computer Science, Washington State University, Pullman, USA. He is currently the director of the school. His current research interests are novel interconnect architectures for manycore chips, on-chip wireless communication networks, and heterogeneous architectures. Dr. Pande currently serves as the Associate Editor-in-Chief (A-EIC) of IEEE Design and Test (D&T). He is on the editorial boards of IEEE Transactions on VLSI (TVLSI) and ACM Journal of Emerging Technologies in Computing Systems (JETC) and IEEE Embedded Systems letters. He was the technical program committee chair of IEEE/ACM Network-on-Chip Symposium 2015 and CASES (2019-2020). He also serves on the program committees of many reputed international conferences. He has won the NSF CAREER award in 2009. He is the winner of the Anjan Bose outstanding researcher award from the college of engineering, Washington State University in 2013.